\documentclass[%
showpacs,preprintnumbers,
nobibnotes,
 amsmath,amssymb,
 aps,
 pre,
 longbibliography,
 lengthcheck,%
]{revtex4-1}

\usepackage{graphicx}
\usepackage{dcolumn}
\usepackage{bm}
\usepackage{hyperref}

\begin{document}

\title{Resonant Response in Non-equilibrium Steady States.}

\author{R. Salgado-Garc\'{\i}a} 
\email{raulsg@uaem.mx}
\affiliation{Facultad de Ciencias, Universidad Aut\'onoma del Estado de Morelos. Avenida Universidad 1001, Colonia Chamilpa, 62209, Cuernavaca Morelos, Mexico.} 
\date{\today} 

\begin{abstract}
The time-dependent probability density function of the order parameter of a system evolving towards a stationary state exhibits an oscillatory behavior if the eigenvalues of the corresponding evolution operator are complex. The frequencies $\omega_n$, with which the system reaches its stationary state, correspond to the imaginary part of such eigenvalues. If the system (at the stationary state) is further driven by a small and oscillating perturbation with a given frequency $\omega$, we formally prove that the linear response to the probability density function is enhanced when $\omega=\omega_n$ for $n\in\mathbb{N}$. We prove that the occurrence of this phenomenon is characteristic of systems that are in a non-equilibrium stationary state. In particular we obtain an explicit formula for the frequency-dependent mobility in terms of the relaxation  to the stationary state of the (unperturbed) probability current. We test all these predictions by means of numerical simulations considering an ensemble of non-interacting overdamped particles on a tilted periodic potential. 
\end{abstract}


\maketitle


\section{INTRODUCTION}


The linear response theory has long been investigated for systems at thermodynamic equilibrium~\cite{einstein,nyquist,onsager1,green1,kubo2,marconia}. Recently, much effort has been made to develop a parallel theory of linear response for systems far from equilibrium and in particular, for those which attain a non-equilibrium stationary state (NESS)~\cite{marconia}. 
Moreover, the study of relaxation phenomena towards a NESS has also become an important issue in a variety of systems. Examples of the latter are  biochemical reaction systems~\cite{quian2010,ge2011non}, growth process~\cite{chou2009changing} and systems in which there occur transitions between stationary states~\cite{asai2011}, among others.

In this work we are concerned with the response to a time-dependent perturbation of systems with (continuous or discrete) markovian dynamics. We report a resonant behavior which is exhibited by systems in a NESS, and a relationship  between such a response function and the relaxation properties of the system towards the NESS.  The occurrence of the resonances in the response function can be intuitively explained as follows. The probability density function (PDF) relaxes to the stationary state through a superposition of oscillations with frequencies $\{ \omega_n \}$. The latter are given by the imaginary part of the (complex) eigenvalues of a given generator $\mathcal{L}$. If this system is further perturbed with an external forcing oscillating with a given frequency $\omega$, then we would expect to have an enhancement of the response to the PDF just at $\omega = \omega_n$. Then, this effect can be perceived in some observable quantities like, for example the probability current. That the system must attain a NESS is a necessary (but not a sufficient) condition for the resonant response to occur. Indeed, it is known that for discrete markovian systems the detailed balance (DB) condition implies that the generator $\mathcal{L}$ can be transformed into a symmetric one $\mathcal{L}^*$~\cite{levin2009markov}. This is then equivalent to say that DB imply real eigenvalues.  Nevertheless, for continuous systems it has been derived a symmetry property~\cite{kurchan1998fluctuation} from the detailed balance, which we show leads to a similar result. This excludes the possibility of having resonant response for markovian systems fulfilling detailed balance. 
 
At this point it is worth to speak about some other works related to the ours. First of all we should point out that the resonance phenomenon in the susceptibility function was first predicted by Ruelle in~\cite{ruelle1,ruelle2} for uniformly and non-uniformly hyperbolic dynamical systems. The former systems are known to have the property that time averages are uniquely determined by a Sinai-Ruelle-Bowen measure. The latter kind of measures have been proposed as candidates for non-equilibrium stationary states~\cite{ruelle3} and some important consequences have been inferred from such hypothesis. An  example of the latter is the Gallavotti-Cohen fluctuation theorem~\cite{gallavotti-cohen1,gallavotti-cohen2}. In the context of dynamical systems, Ruelle stated that the susceptibility function should have two types of singularities in the $\omega$-complex plane: some of them related to the unstable dynamics and others corresponding to the stable one~\cite{ruelle1,ruelle2} of the hyperbolic dynamics. Such singularities are called resonances (see Ref.~\cite{ruelle3} section 4.9 and~\cite{ruelle2}). In particular Ruelle stated that the singularities corresponding to the stable dynamics ``would correspond to resonances in the `oscillations of the system around its attractor' ''~\cite{ruelle2}, 
while the singularities corresponding to the unstable one are the same as those of the spectral density (i.e., the fluctuation-dissipation relation holds only for the unstable contribution~\cite{ruelle1,ruelle2}). 
The general framework of Ruelle is applicable to chaotic systems and the presence of such resonances has been verified analytically and numerically~\cite{cessac2004stable,cessac2007linear,cessac2007does} in specific chaotic models. However, in the context of stochastic systems attaining a non-equilibrium stationary state such phenomena has not been previously studied (up to the knowledge of the author). In this paper we  extend part of the Ruelle's ideas to the stochastic domain.  
It is worth mentioning that the linear response for the case of stochastic systems is very different from that of the chaotic case~\cite{marconia}. For example, difficulties related with the possibility that the Sinai-Ruelle-Bowen measures be singular (i.e. not absolutely continuous with respect to Lesbegue) and the decomposition of the state space into stable and unstable directions, are absent in our case. Thus, the phenomena observed in the linear response for the models analyzed in~\cite{cessac2004stable,cessac2007linear,cessac2007does} are different and more complex than the behavior found in the systems explored in this work.

 This paper is organized as follows, in Section~\ref{sec:theory} we prove that the resonance in the linear response may occur if the system reaches a NESS via oscillations due to complex eigenvalues. In Section~\ref{sec:tilted-potentials} we show the occurrence of the resonant phenomenon in titled periodic potentials by means of numerical experiments. Finally in Section~\ref{sec:conclusion} we give our conclusions.  Some appendices are included containing detailed calculations.


\section{THE LINEAR RESPONSE AND THE RELAXATION OF THE PDF}
\label{sec:theory}


Consider an unperturbed markovian system with state space $X$, whose dynamics is defined by a generator $\mathcal{L}$. Throughout this work we will use the Dirac's bracket notation as in Ref.~\cite{kurchan1998fluctuation}. We denote the time-dependent PDF of such a system by $\rho(x,t) = \langle x | \rho (t) \rangle $, with $x\in X$. Then, the evolution equation for $|\rho \rangle $ can be written as,
\begin{equation}
\frac{\partial |\rho \rangle}{\partial t} = \mathcal{L} | \rho \rangle.
\label{eq:unperturbed}
\end{equation} 
By using the separation ansatz  $| \rho(t) \rangle = e^{\lambda t}  |P_\lambda \rangle$, we arrive at the eigenvalue problem for $\mathcal{L}$,
\begin{equation}
\mathcal{L} | P_\lambda  \rangle = \lambda  |P_\lambda \rangle.
\end{equation}
It is known that the largest eigenvalue is $\lambda = 0$. Assuming non-degeneracy of the latter, the corresponding eigenvector $ |P_0 \rangle $ is the unique stationary solution of~\eqref{eq:unperturbed} on the set $\Delta_X $ of all normalized PDFs. The set $\Delta_X $ of probability densities is commonly called the \emph{simplex space} and is defined as $\Delta_X:= \{ |\rho \rangle : \langle - |\rho \rangle = 1\}$. Here $|-\rangle$ is a probability density (called the \emph{flat  distribution}) perpendicular to the simplex $\Delta_X$  and is defined by the equation $\langle-| x\rangle = 1$. It is clear that all the other eigenvalues satisfy $\mbox{Re}(\lambda) < 0$. Moreover, the eigenvectors corresponding to the non-vanishing eigenvalues are parallel to the simplex $\Delta_X $, i.e.,  $\langle - | P_\lambda \rangle  = 0$ for all $\lambda \not=0$. 

We now consider the perturbed  evolution equation,
\begin{equation}
\frac{\partial | \rho_p \rangle }{\partial t} = \mathcal{L} |\rho_p\rangle + \varepsilon e^{i \omega t} \mathcal{L}_p |\rho_p \rangle,
\label{eq:perturbed}
\end{equation}
where $\varepsilon >0$ is a small control parameter, $\omega >0$ is a given frequency and $\mathcal{L}_p$ is a perturbing operator. We now follow the usual way to obtain the linear response to the PDF~\cite{marconia}. First we assume that the perturbed solution can be written as 
\begin{equation}
|\rho_p \rangle \approx |P_0 \rangle + \varepsilon e^{i \omega t} |R \rangle.
\label{eq:ansatz-pdf}
\end{equation}
In the above ansatz, the deviation from the stationary state of the perturbed system, with respect to the unperturbed one, is represented by $\varepsilon |R \rangle e^{i \omega t} $. As $ |\rho_p \rangle$ is a normalized probability vector, it follows that $|R \rangle$ is a vector parallel to the simplex  $\Delta_X$, i.e., $\langle -| R \rangle = 0$. We will refer to $|R \rangle$ as the \emph{response to the stationary PDF}. It is clear that in the ansatz~\eqref{eq:ansatz-pdf} we use the basic assumption that the perturbed PDF will eventually be periodic, oscillating with the same frequency of the perturbation. Using Eq.~\eqref{eq:ansatz-pdf} in~\eqref{eq:perturbed} we obtain the following equation for $|R \rangle$,
\begin{equation}
\left( i \omega  - \mathcal{L} \right) |R \rangle =  \mathcal{L}_p |P_0 \rangle.
\label{eq:response}
\end{equation}
The linear response to the PDF has the formal solution $|R \rangle = \left( i \omega  - \mathcal{L} \right)^{-1} \mathcal{L}_p |P_0 \rangle$.  We now use the fact that the inverse operator $\left( s  - \mathcal{L} \right)^{-1} $ can be formally written in an integral representation as~\cite{evans2008statistical},
\[
\left( s  - \mathcal{L} \right)^{-1}  = \int_0^\infty dt e^{-st}e^{\mathcal{L}t},
\]
for $\mbox{Re}(s)>0$. If we set $s=\delta - i\omega$, with $\delta >0$, we can write $R$ as,
\begin{equation}
|R\rangle = \lim_{\delta \to 0^+} \int_0^\infty dt e^{-st}e^{\mathcal{L}t} \mathcal{L}_p |P_0\rangle.
\label{eq:R-integral-representation}
\end{equation}
At this point it is necessary to point out that the perturbing operator is such that  $ \langle -|\mathcal{L}_p = 0$. The latter follows from the probability conservation of both, the perturbed and the unperturbed system~\cite{kurchan1998fluctuation}. Such a property lets us to interpret $e^{\mathcal{L}t} \mathcal{L}_p | P_0 \rangle$ as the time-evolution of a vector parallel to the simplex with initial condition $\mathcal{L}_p | P_0 \rangle$. With this in mind we can see that $e^{\mathcal{L}t} \mathcal{L}_p P_0$ can be expanded in a series of the eigenvectors of $\mathcal{L}$ as,
\begin{equation}
e^{\mathcal{L}t} \mathcal{L}_p | P_0 \rangle = \sum_{\lambda \not= 0} c_\lambda  e^{\lambda t} | P_\lambda \rangle.
\label{eq:expansion}
\end{equation}
Notice that the eigenvector corresponding to $\lambda = 0$ is excluded, since $e^{\mathcal{L}t} \mathcal{L}_p | P_0 \rangle$  has no components perpendicular to the simplex. If we put the last expression into~\eqref{eq:R-integral-representation} we obtain,
\begin{eqnarray}
| R \rangle &=&  \lim_{\delta \to 0}  \sum_{\lambda \not= 0} \int_0^\infty dt e^{-(s-\lambda)t}c_\lambda  | P_\lambda, \rangle
\nonumber
\\
&=&  \lim_{\delta \to 0}  \sum_{\lambda \not= 0} \frac{c_\lambda | P_\lambda \rangle}{s-\lambda},
\nonumber
\end{eqnarray}
which results in,
\begin{equation}
| R \rangle = \sum_{\lambda \not= 0} \frac{c_\lambda  | P_\lambda \rangle}{i\omega - \lambda}.
\label{eq:R-resonant}
\end{equation}
The mechanism leading to the resonant response becomes clear from Eq.~\eqref{eq:R-resonant}: the enhancement of the amplitude of the response to the PDF, as a function of $\omega$, is a consequence of the oscillations of the PDF in the simplex space towards its NESS. At this point it is important to stress that under our assumptions, \emph{ the resonant behavior predicted by Eq.~\eqref{eq:R-resonant} is a characteristic of non-equilibrium stationary states which is not present in systems at equilibrium}. This follows from the fact that DB implies that the eigenvalues of $\mathcal{L}$ are real. The latter is a consequence of the fact that DB can be used to transform the operator $\mathcal{L}$ in a symmetric one. This has been shown in Ref.~\cite{levin2009markov} if $\mathcal{L}$ is a transition matrix with a discrete state space $X$. In Appendix~\ref{ape:DetBal-RealEigenVals} we give a proof, based on a symmetry property derived from the DB by Kurchan~\cite{kurchan1998fluctuation}, for the equivalent statement when $\mathcal{L}$ stands for a Fokker-Planck operator.

%
%

The above-described resonant behavior occurs at the level of probability densities since it corresponds to an enhancement of the PDF oscillations. However, in a real experiment it is not always possible to measure the PDF itself. Then, it is important to explore how this behavior is displayed by a given observable. Here we analyze the  case in which such an observable is the probability current. Consider the perturbed system~\eqref{eq:perturbed} in the asymptotic (time-dependent) state $|\rho_p \rangle$ and  let $\mathcal{K}(t)$ be the operator for the probability current. The expected value of the operator $\mathcal{K}(-\infty)$ gives the probability current at the stationary state,
\[ 
j_0 = \langle - |\mathcal{K}(-\infty) |P_0 \rangle. 
\]
In a number of cases the generator $\mathcal{L}$ can be factorized as $\mathcal{L} =\mathcal{D} \mathcal{K}_0$. For example, if $\mathcal{L}$ represents the Fokker-Planck operator, $\mathcal{D} = \partial_x$ and $\mathcal{K}_0= f - \beta^{-1}\partial_x$, with $f$ as minus the gradient of a given potential $V$. In a similar way we  also assume that the perturbing operator $\mathcal{L}_p$ can be factorized as $\mathcal{D} \mathcal{K}_p$, and therefore, the probability current operator can be written as $ \mathcal{K}(t) = \mathcal{K}_0 + \varepsilon \mathcal{K}_p e^{i\omega t} H(t)$. Here $H(t)$ stands for the heavyside function. From the form of $K(t)$ we infer that $\mathcal{K}(-\infty) = \mathcal{K}_0$.  The probability current for the perturbed system $j_p$ can be obtained by,
\[
j_p(t) = \langle -|(\mathcal{K}_0 + \varepsilon \mathcal{K}_p e^{i\omega t} H(t) )| \rho_p(t) \rangle .
\]
With the above expression we can calculate the probability current at first order in $\varepsilon$. Using Eq.~\eqref{eq:ansatz-pdf} we obtain for $t>0$,
\[
j_p(t) = j_0 + \varepsilon e^{i \omega t} \left( \langle -| \mathcal{K}_p |P_0\rangle  +  \langle -| \mathcal{K}_0 |R(\omega)\rangle \right),
\]
and from the last expression we identify
\begin{equation}
\label{eq:mu_omega}
\mu(\omega) := \langle -|\mathcal{K}_p  | P_0\rangle +  \langle -| \mathcal{K}_0 | R(\omega) \rangle , 
\end{equation}
with the linear response to the probability current (the frequency dependent mobility). Substituting $|R\rangle $ by the expression given by Eq.~\eqref{eq:R-integral-representation} into the above equation we obtain
\[
\mu(\omega)=  \langle - | \mathcal{K}_p | P_0\rangle  + \lim_{\delta \to 0} \int_0^\infty dt e^{-st}   \langle - | \mathcal{K}_0 e^{\mathcal{L}t} \mathcal{L}_p | P_0 \rangle,
\]
or equivalently,
\begin{equation}
\label{eq:mu:middle}
\mu(\omega)=  \langle - | \mathcal{K}_p | P_0\rangle  + \int_0^\infty dt e^{i\omega t}   \langle - | \mathcal{K}_0 e^{\mathcal{L}t} \mathcal{L}_p | P_0 \rangle.
\end{equation}
In order to give some physical interpretation to the above expression lets consider again the evolution equation~\eqref{eq:unperturbed} of the unperturbed system. First, we prepare the system with the initial PDF $| \rho^*(0) \rangle  := |P_0 \rangle + \gamma \mathcal{L}_p |P_0\rangle $, where $\gamma$ is, in some extent, arbitrarily chosen, just requiring that $ | \rho^*(0)\rangle > 0$. Here $ |P_0\rangle$ represents the stationary state of our system and $\gamma \mathcal{L}_p | P_0\rangle$ is a vector parallel to the simplex. This observation makes it clear that $|\rho^*(0) \rangle$ fulfill the normalization condition. The formal expression for the time-dependent PDF, with the described initial condition, is given by $| \rho^*(t) \rangle = |P_0 \rangle  + e^{\mathcal{L}t} \mathcal{L}_p |P_0 \rangle$. If we use the latter to calculate the probability current $j^*(t)$ for this specific system we found that, 
\begin{equation}
j^*(t) = j_0 + \gamma \langle -|  \mathcal{K}_0 e^{\mathcal{L}t} \mathcal{L}_p | P_0 \rangle.
\end{equation}
The last results lets us rewrite Eq.~\eqref{eq:mu:middle} as,
\begin{equation}
\label{eq:RRrelation}
\mu(\omega)=  \langle - |\mathcal{K}_p |P_0\rangle +  \gamma^{-1} \int_0^\infty dt e^{i\omega t} ( j^*(t) - j_0),
\end{equation}
which relates the relaxation towards the stationary state with the linear response to an external perturbation.


\section{RESONANT RESPONSE IN TILTED PERIODIC POTENTIALS: NUMERICAL EXPERIMENTS}
\label{sec:tilted-potentials}


In this section we show the presence of the resonant behavior predicted by Eq.~\eqref{eq:R-resonant} in tilted periodic potentials by means of numerical simulations. In particular we test the relationship between the relaxation to the steady state and the response to an external perturbation given by Eq.~\eqref{eq:RRrelation}. First we consider an unperturbed system consisting of non-interacting overdamped particles in a tilted periodic potential $V(x)$ with period $L$ and tilt $F_0$ (i.e. $V(x+L) = V(x) - F_0 L$) subjected to a gaussian white noise. The stochastic dynamics of these particles is ruled by the Langevin equation,
\begin{equation}
\tilde \gamma \frac{dx}{dt} = f(x)+\xi(t),
\label{eq:langevin-non-pert}
\end{equation}
where $\tilde \gamma$ is the friction coefficient which, as usual, will be set to one. The function $f(x)$ is minus the gradient of the tilted periodic potential $V(x)$. With this we have that $f(x)$ is periodic with period $L$. The term $\xi(t)$ represents a gaussian white noise with zero mean and correlation $\langle \xi(0)\xi(t) \rangle = 2\beta^{-1} \delta(t)$. Here $\beta$ is the inverse temperature of the system. 

The Fokker-Planck equation corresponding to the Langevin dynamics~\eqref{eq:langevin-non-pert} is given by,
\begin{equation}
\frac{\partial \rho }{\partial t} = \mathcal{L}_{\mathrm{FP}} \rho,
\label{eq:FP-unperturbed}
\end{equation}
where $\mathcal{L}_{\mathrm{FP}}$ is given by
\begin{equation}
\mathcal{L}_{\mathrm{FP}}  = - \frac{\partial }{\partial x} \bigg(  f(x) \ - \beta^{-1} \frac{\partial \  }{\partial x} \bigg).
\label{eq:FP-unperturbed-operator}
\end{equation}
Unfortunately the exact eigenvalues $\lambda_n$ for the Fokker-Planck operator $\mathcal{L}_{\mathrm{FP}}$ are not generally known. Nevertheless, the author and coworkers have given an approximation to $\lambda_n$ in~\cite{salgado-garcia}. Such an approximation is valid for large values of the tilt, $F_0 \gg \max\{f(x)\} - \min\{ f(x)\}$, and is given by $\lambda_n = i2\pi n j_0/L + (2\pi n/L)^2 D_{\mathrm{eff}}$. Here $j_0$ is the probability current, $D_{\mathrm{eff} }$ is the effective diffusion coefficient, $L$ is the period of the (tilted) periodic potential and $n\in \mathbb{N}$. From this it follows that $\omega_n = 2\pi n j_0/L $ is an approximation to the natural frequencies of the system.

%
\begin{figure}[t]
\begin{center}
\scalebox{0.45}{\includegraphics{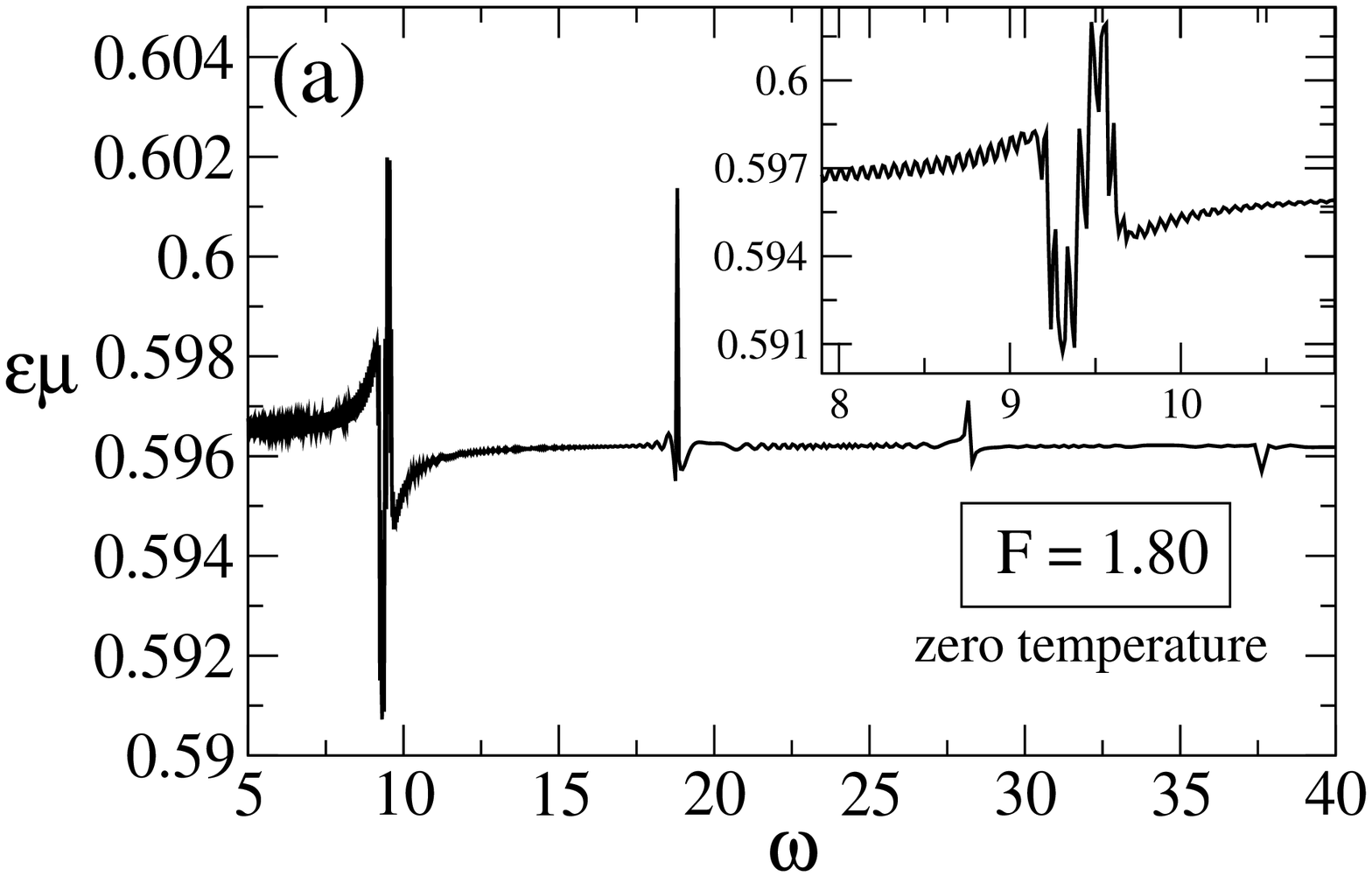}}
\\
\scalebox{0.45}{\includegraphics{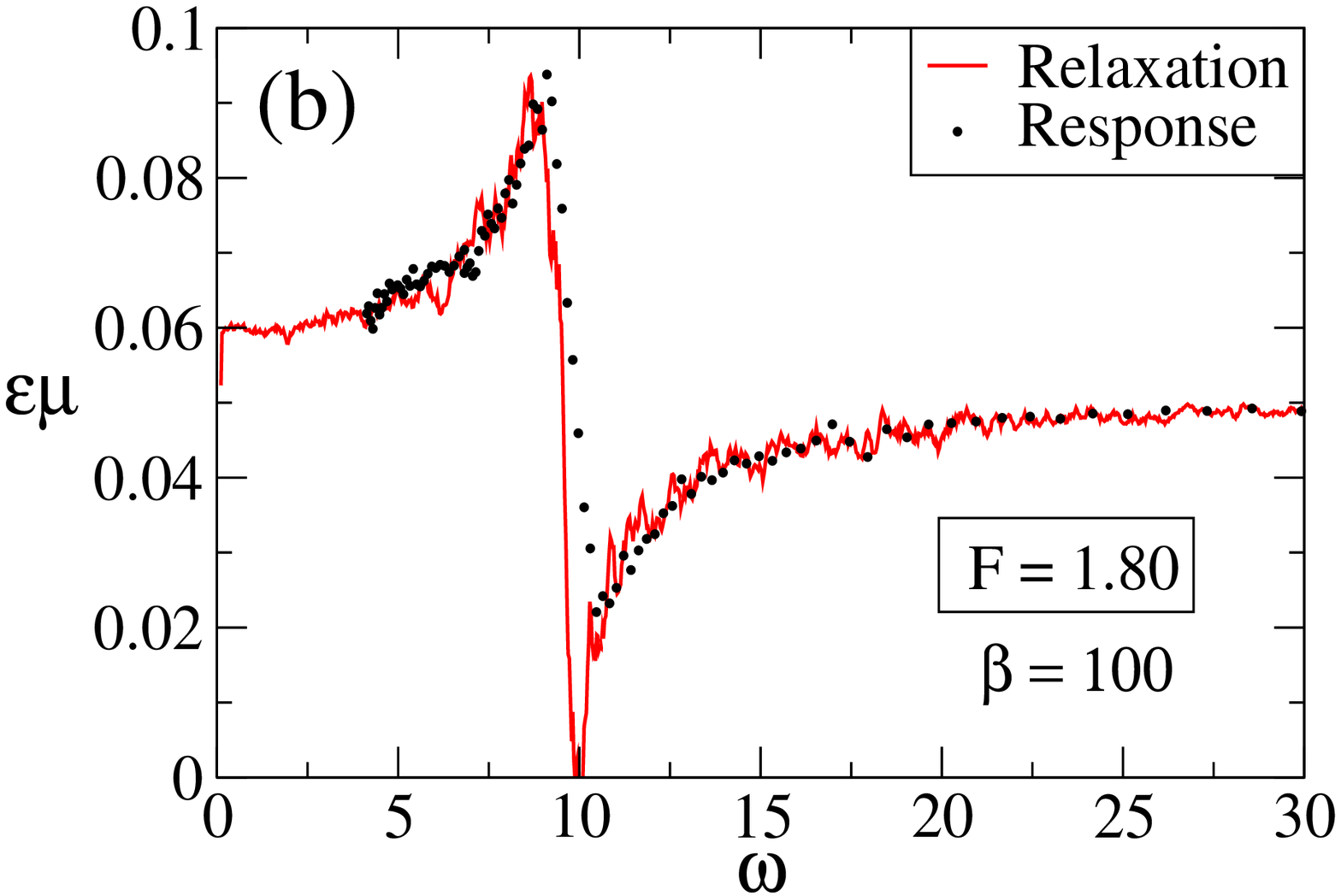}}
\end{center}
     \caption{(Color online) 
     Frequency-dependent mobility $\mu(\omega)$ for the tilted periodic potential $V(x) = \cos(2\pi x)/(2\pi) -x F_0 $ with $F_0  = 1.80$ (a) at zero temperature and (b) at a temperature $\beta^{-1}= 0.01$. The external perturbation is taken as $0.05 \times \cos(\omega t)$ (i.e. $\varepsilon = 0.050$). In (a) we can see that the deterministic response exhibits several peaks corresponding to the resonance frequencies $\omega_n = 2\pi n \bar{v} /L$ (see text). The inset shows an amplification in the region near the main resonance peak occurring at $\omega = 9.4$ to better appreciate the complex structure of the response. In (b) we observe a similar behavior of the response vs. frequency at a finite temperature. However, most of the resonant peaks observed in the deterministic case disappear. We also see that the relaxation of the probability current obtained by~\eqref{eq:relaxation-formula} (solid line), and the response to the stationary current (filled circles) are related according to the expression~\eqref{eq:RRrelation} (see text for details).  
     }
\label{fig:Det_B100_F1.80}
\end{figure}
%

%
%
\begin{figure}[t]
\begin{center}
\scalebox{0.45}{\includegraphics{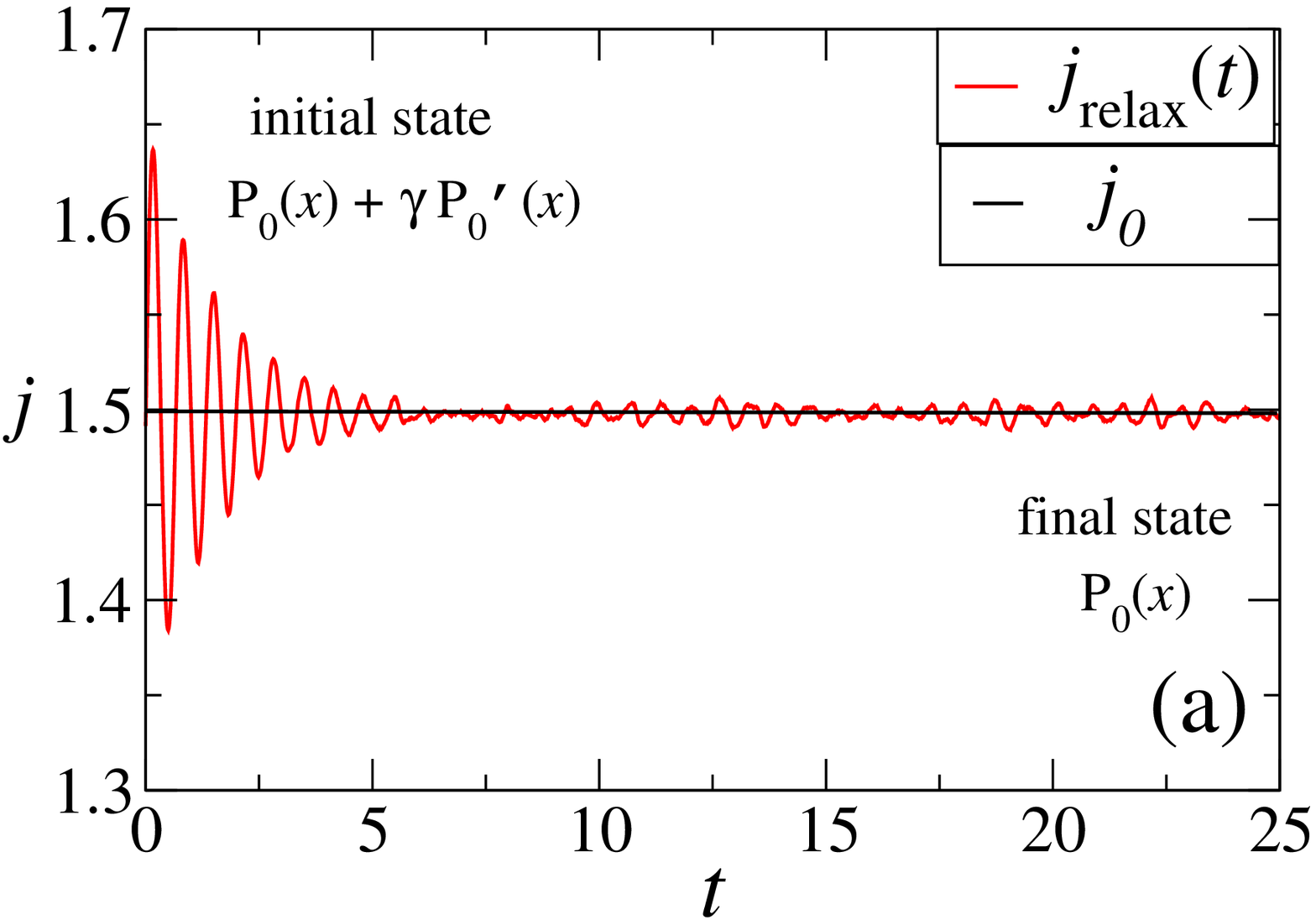}}
\\
\scalebox{0.45}{\includegraphics{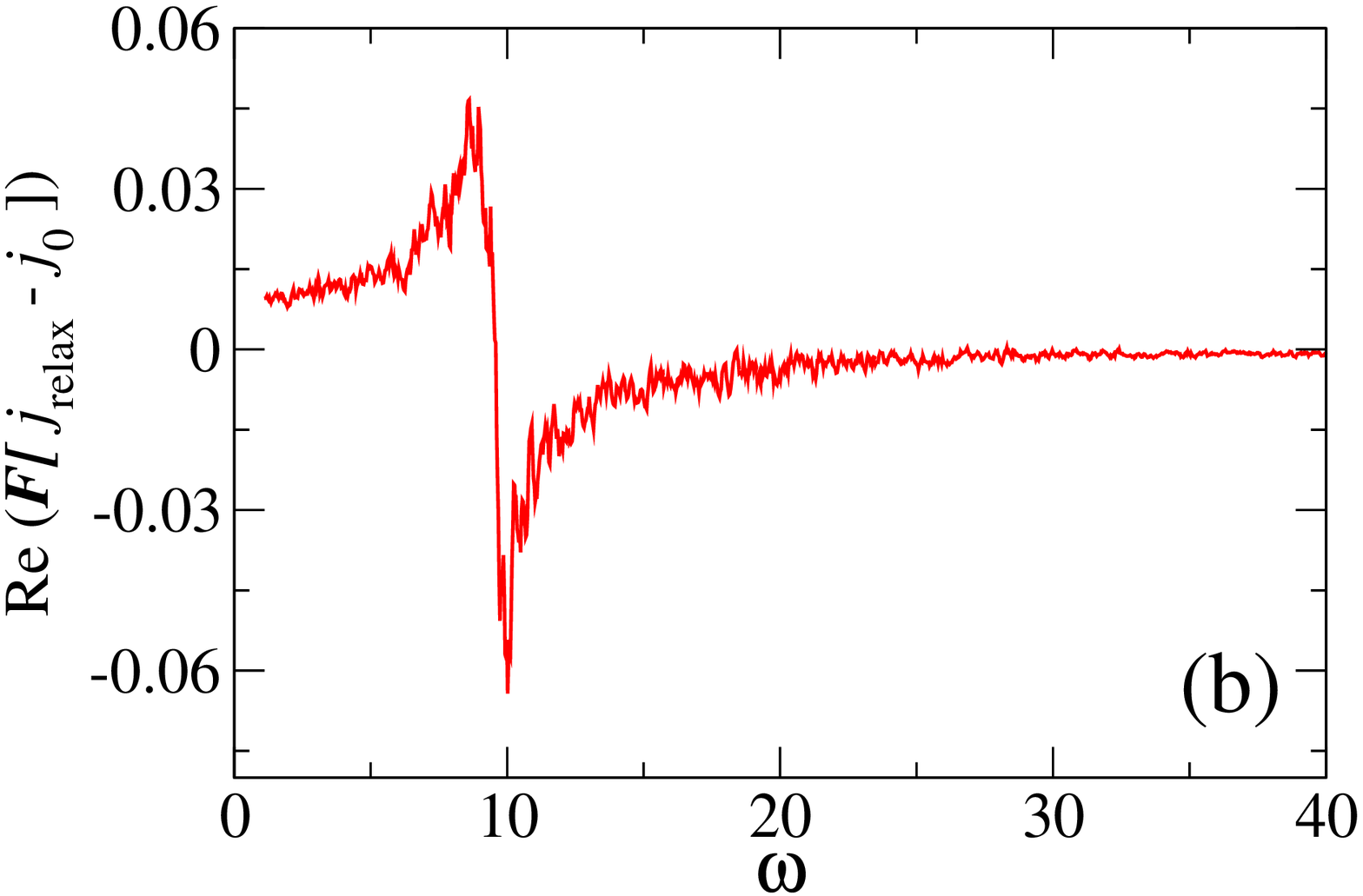}}
\end{center}
     \caption{(Color online) 
     The relaxation of the probability current towards the stationary state. Figure (a) shows the probability current $j_{relax}$ as a function of time for the unperturbed system~\eqref{eq:langevin-non-pert} with $V(x) = \cos(2\pi x)/(2\pi) -x F_0 $,  $F_0  = 1.80$ and $\beta^{-1}= 0.01$. We obtained this time series through a simulation of $20000$ particles obeying the Langevin equation~\eqref{eq:langevin-non-pert} up to a time $t=100$ arb. units. The particles were initially distributed according to the PDF $P_0(x)+ \gamma P^\prime (x)$ with $\gamma  =0.080$. 
     (b) The real part of the Fourier transform of $j_{\mathrm{relax}} - j_0$. To calculate the Fourier transform we used a time series for the probability current shown in (a). After that we used formula~\eqref{eq:relaxation-formula} to obtain the ``relaxation curve'' shown in Fig.~\ref{fig:Det_B100_F1.80}b (solid line).  We proceeded in the same way to calculate the  ``relaxation curves'' shown in the subsequent figures.
     }
\label{fig:jrelax_FT_F1.80}
\end{figure}
%
%
%
%

According to Section~\ref{sec:theory}, we have that if this system is perturbed by a time dependent periodic forcing $\varepsilon F(t)$, characterized by a frequency $\omega$, then, a resonant behavior in the PDF will be exhibited at some $\omega = \omega^*$. Such a perturbed system is modeled by the Langevin equation,
\begin{equation}
\tilde \gamma \frac{dx}{dt} = f(x)+\varepsilon F(t) + \xi(t),
\label{eq:langevin-perturbed}
\end{equation}
with $\tilde \gamma $ set to one. To achieve the numerical experiments showing the occurrence of resonances at the frequencies given above,  we chose the tilted periodic potential as $V(x) = \cos(2\pi x)/(2\pi) -x F_0 $. With this potential we have that $f(x) = \sin(2\pi x) + F_0$ and that $L=1$. The external forcing $F(t)$ is taken as $F(t) = \cos(\omega t)$ and $\varepsilon$, the small control parameter, is fixed to $\varepsilon = 0.05$. 

In Fig.~\ref{fig:Det_B100_F1.80}a we show the mobility (the amplitude of the current oscillations with respect to the stationary current) as a function of the frequency for the deterministic tilted periodic potential with tilt $F_0 = 1.80$. In such a figure we can observe that at zero temperature the system exhibit several resonant peaks at the frequencies given by the formula $\omega_n = 2 \pi n \bar{v}/L$ (see Eq.~ \eqref{eq:freq-det} from Appendix~\ref{ape:zero-temp}). Here the mean velocity is $\bar{v} \approx 1.5$, and therefore $\omega_n \approx 9.4 n$. This behavior can be readily appreciated in Fig.~\ref{fig:Det_B100_F1.80}a.  In Fig.~\ref{fig:Det_B100_F1.80}b we show the behavior for the system at a finite temperature chosen as $\beta^{-1} = 0.01$. In this figure we see that frequency-dependent mobility $\mu(\omega)$ (filled circles) displays a behavior similar to that of  its deterministic counterpart. It exhibits a resonant peak at the natural frequency (which is estimated, according to Ref~\cite{salgado-garcia} as $\omega_0 = 2\pi j_0/L \approx 9.4$). However, the successive resonant peaks seems to be suppressed by the noise in the thermalized system.

%
\begin{figure}[t]
\begin{center}
\scalebox{0.435}{\includegraphics{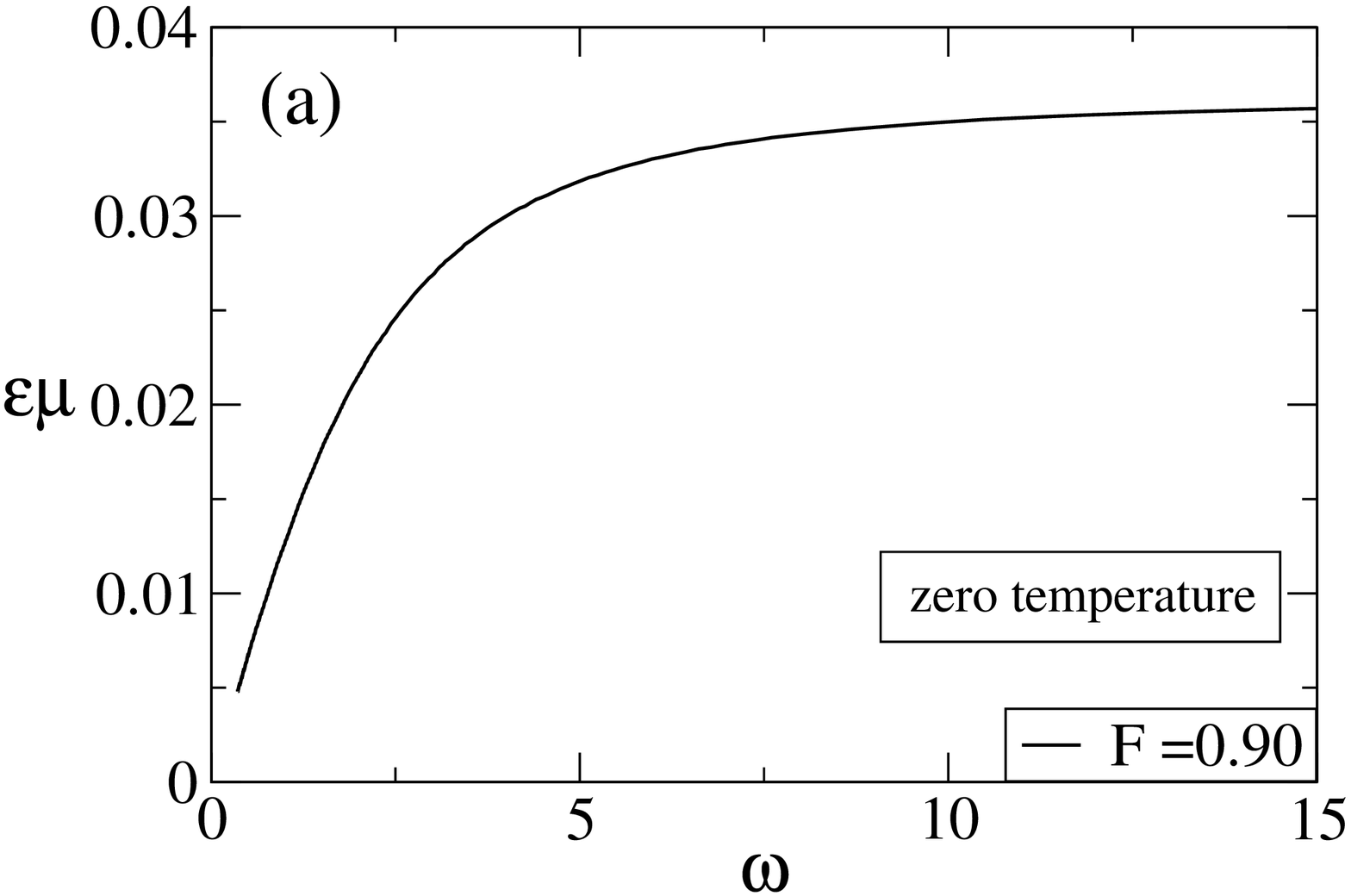}}
\\
\scalebox{0.435}{\includegraphics{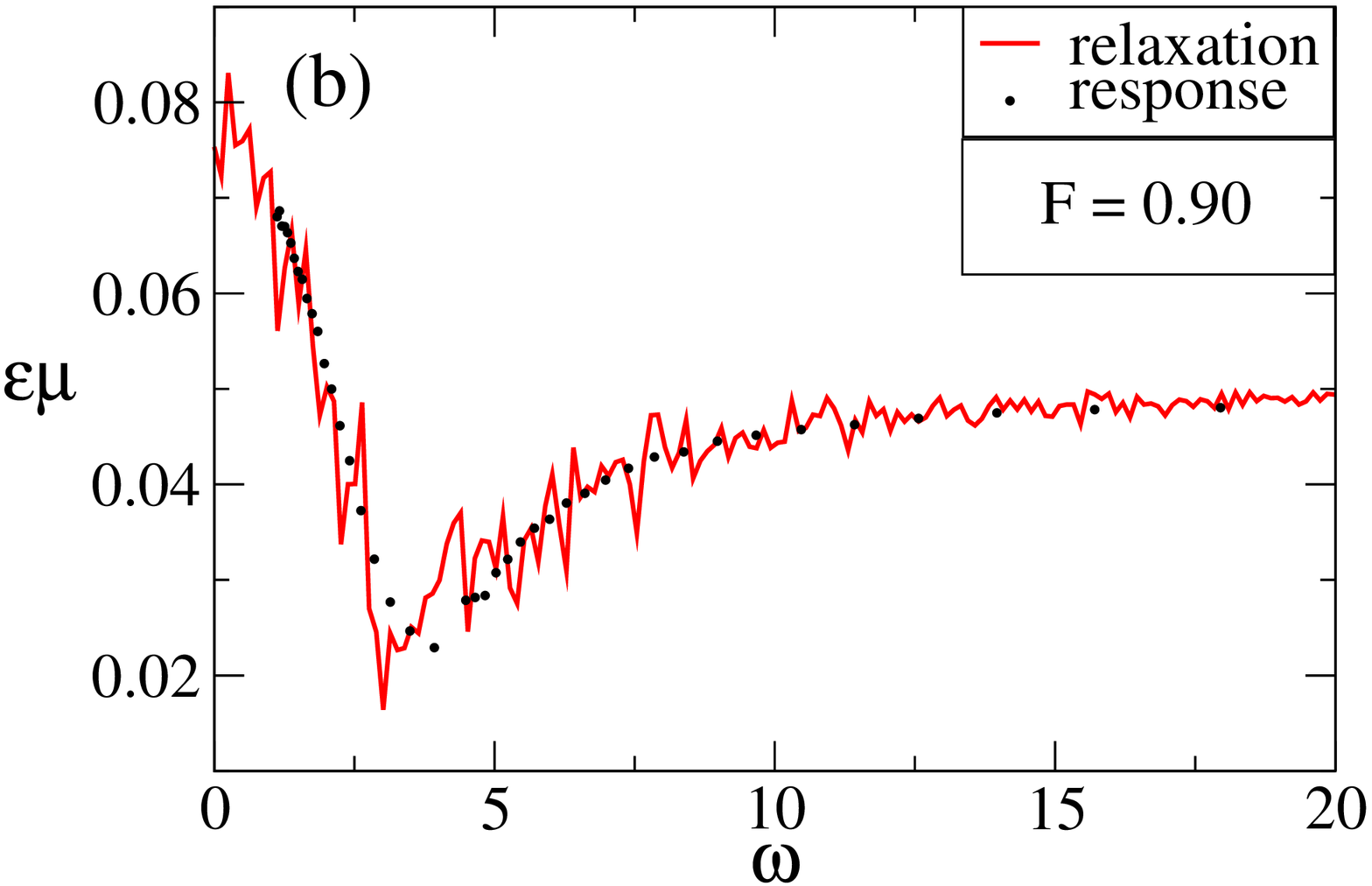}}
\end{center}
     \caption{(Color online)
     Frequency-dependent mobility $\mu(\omega)$ for the tilted periodic potential $V(x) = \cos(2\pi x)/(2\pi) -x F_0 $ with $F_0  = 0.90$ (a) at zero temperature and (b) at a temperature $\beta^{-1}= 0.01$. The external perturbation is taken as $0.05 \times \cos(\omega t)$ (i.e. $\varepsilon = 0.050$). In (a) we can see that the deterministic response is monotonically decreasing curve. In (b) we observe a non-monotonous behavior of the linear response (filled dots), which is quite different from its deterministic counterpart. We also compare the relaxation of the current given by~\eqref{eq:relaxation-formula} (solid line) with the linear response (filled circles) in order to test the relaxation-response relation given by~\eqref{eq:RRrelation}.
     }
\label{fig:Det-B100F0.90}
\end{figure}
%

%
%
\begin{figure}[t]
\begin{center}
\scalebox{0.45}{\includegraphics{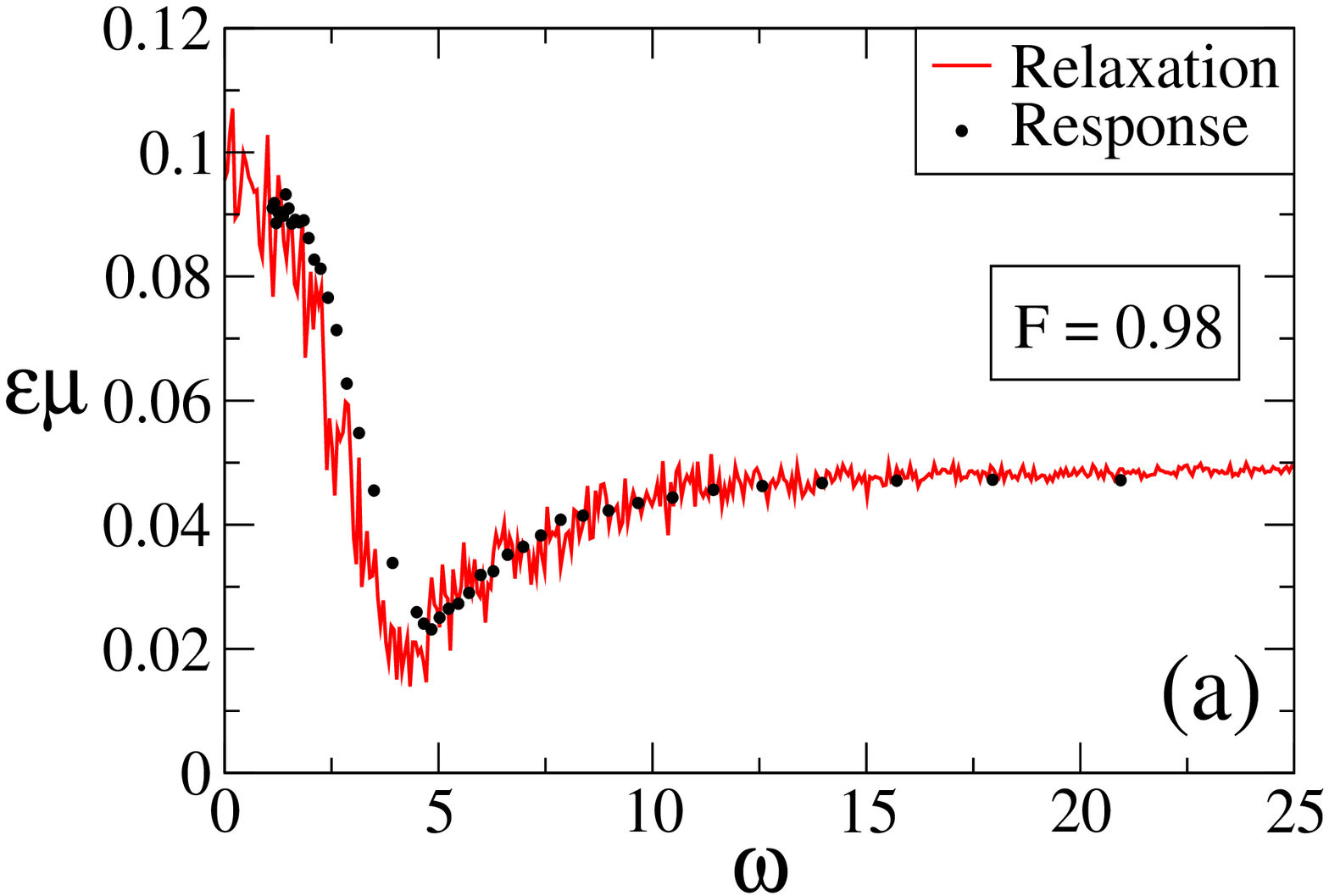}}
\\
\scalebox{0.45}{\includegraphics{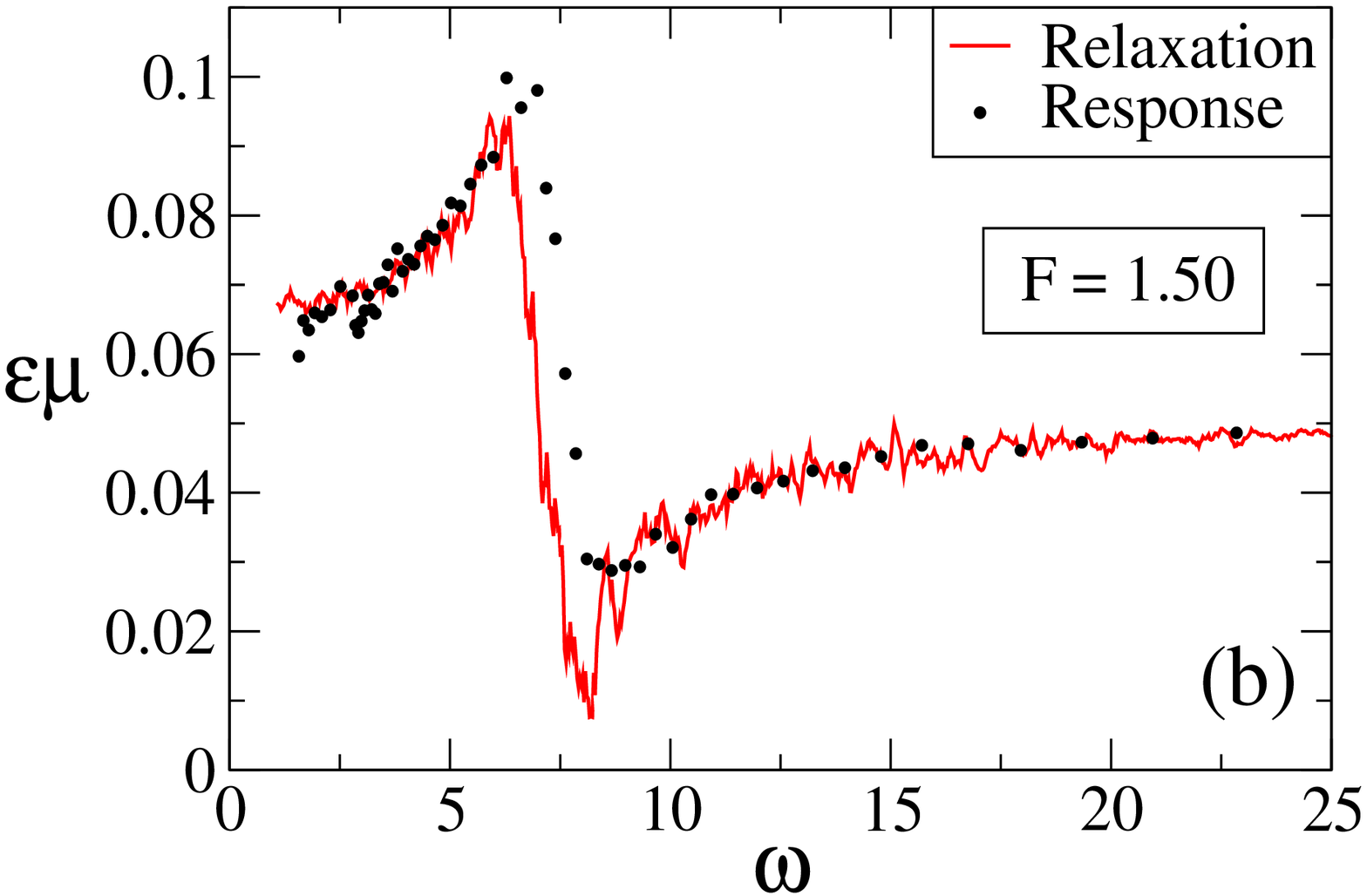}}
\end{center}
     \caption{(Color online)
     Frequency-dependent mobility $\mu(\omega)$ for the tilted periodic potential $V(x) = \cos(2\pi x)/(2\pi) -x F_0 $ at temperature $\beta^{-1}= 0.01$ (a) for  $F_0 = 0.98$ (below the critical force $F_c = 1$) (b) for  $F_0 = 1.50$ (above the critical force $F_c = 1$). The external perturbation is taken as $0.05 \times \cos(\omega t)$ (i.e. $\varepsilon = 0.050$).
     The graphs compares the mobility obtained by the response of the system (filled circles) and the one obtained by the relaxation of the probability current (solid line) given by~\eqref{eq:RRrelation}. The latter formula is shown to be right within the accuracy of our numerical simulations.
          }
\label{fig:RR.F0.98.F1.50}
\end{figure}
%
%
%
%

Besides the response to the probability current, in Fig.~\ref{fig:Det_B100_F1.80}b we have also plotted the numerically obtained ``relaxation curve'' (solid line), i.e., the right-hand side of Eq.~\eqref{eq:RRrelation}. Within the accuracy of our simulations we note that a good approximation of the relaxation curve to the response to the probability current at the stationary state given by Eq.~\eqref{eq:RRrelation}.
We now explain how the ``relaxation curve'' was calculated. First we prepared the unperturbed system with the initial PDF given by $P_{\mathrm{initial}} (x) = P_0(x) + \gamma \mathcal{L}_p P_0(x)$ with $\gamma = 0.08$. Here the perturbing operator $\mathcal{L}_p$ is given by $\mathcal{L}_p = -\partial_x $ and $P_0(x)$ is the stationary PDF takes the form,
\[
P_0(x) = \frac{1}{N_0} \exp[-\beta V(x)]\int_{x}^{x+L} \exp[\beta V(y)] dy,
\]  
with $N_0$ a constant such that $\int_0^L P_0(x)dx = 1$. Once we have prepared the system in such state, we let it evolve to the stationary state $P_0(x)$. We then measure the time-dependent probability current $j_{\mathrm{relax}} (t)$ from $t=0$ up to a given time $T_0$ which is large enough to assume that the system has reached the stationary state. 
In Fig.~\ref{fig:jrelax_FT_F1.80}a we plotted  $j_{\mathrm{relax}} (t)$ for the potential $V(x)$ and the parameter values used to perform the numerical experiment corresponding to Fig.~\ref{fig:Det_B100_F1.80}b. As stated by  the right-hand side of the equation~\eqref{eq:RRrelation}, we use the time series obtained above to calculate the real part of the Fourier Transform of $j_{\mathrm{relax}}(t)-j_0$ shown in Fig.~\ref{fig:jrelax_FT_F1.80}b. This quantity is then used to calculate
\begin{equation}
\left\langle - |\mathcal{K}_p| P_0 \right\rangle + \gamma^{-1} \int_0^\infty e^{i \omega t}(j_{\mathrm{relax}}(t) - j_0)dt 
\label{eq:relaxation-formula}
\end{equation}
which corresponds to the ``relaxation curve'' displayed in Fig.~\ref{fig:Det_B100_F1.80}b (solid line). Note that the first term in the above expression equals one since $\mathcal{K}_p = \mbox{Id}$ in this case.

In Fig.~\ref{fig:Det-B100F0.90}a we show the mobility as a function of the frequency for the deterministic tilted periodic potential for a tilt $F_0 = 0.90$. The behavior of the amplitude oscillations in this case is a monotonically decreasing curve whose minimum occurs at $\omega = 0$. This behavior is consistent with the analytically predicted in the Appendix~\ref{ape:zero-temp} for this case. In Fig.~\ref{fig:Det-B100F0.90}b we show the mobility for the very system but at a temperature $\beta^{-1} =0.01$. All the other parameters remain the same as in the case $F=1.80$. In this figure we can see that the response becomes a non-monotonous curve. This is an important feature of the tilted periodic potential below the critical tilt since the noise induces a behavior which is quite different from its deterministic counterpart. This non-monotonous behavior can be seen as emerged from a competition between the deterministic dynamics and the noisy escape rate. On the one hand, at low frequencies we have that the dynamics of the system is dominated mainly by the noise. This is due to the fact that the slow variation of the perturbation lets the particles escape from one potential well to another in every time period. Thus, many particles increase and decrease their velocities, giving large amplitudes to the current oscillations. As the frequency increases, the number of particles escaping from the potential wells during each oscillation diminishes. Then, the amplitude of these current oscillations no longer profits from the escape events and the amplitude of the current starts to behave as in the determinist case. This behavior can be observed in Fig.~\ref{fig:Det-B100F0.90}b: above some frequency value (approximately at $\omega  = 5$) the response curve mimics the form of its deterministic counterpart shown in Fig.~\ref{fig:Det-B100F0.90}a. Figure~\ref{fig:Det-B100F0.90}b also shows that the relation between the relaxation and the response \eqref{eq:RRrelation} also holds for this case within the accuracy of our numerical experiments.

%
%
\begin{figure}[t]
\begin{center}
\scalebox{0.45}{\includegraphics{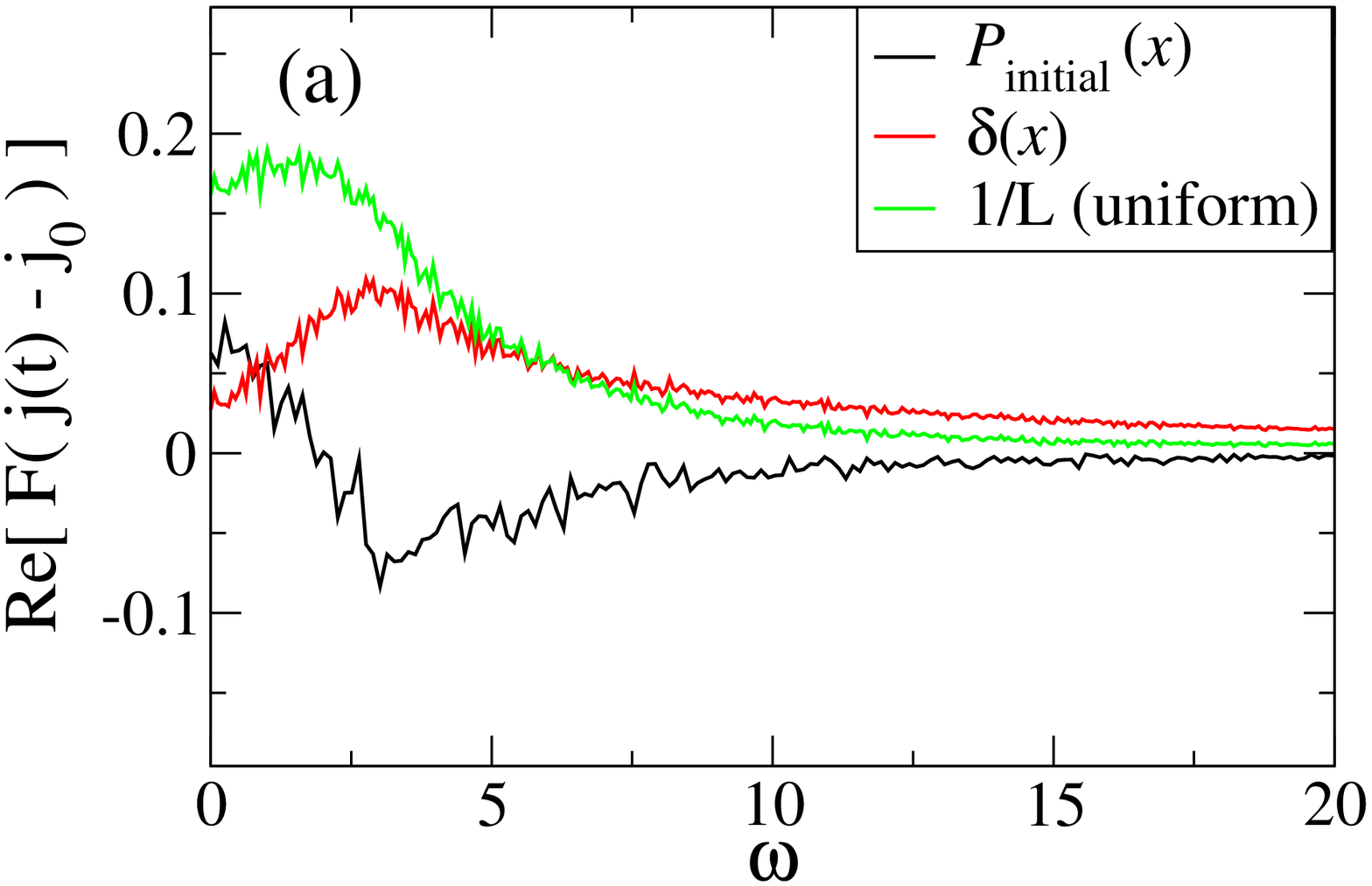}}
\\
\scalebox{0.45}{\includegraphics{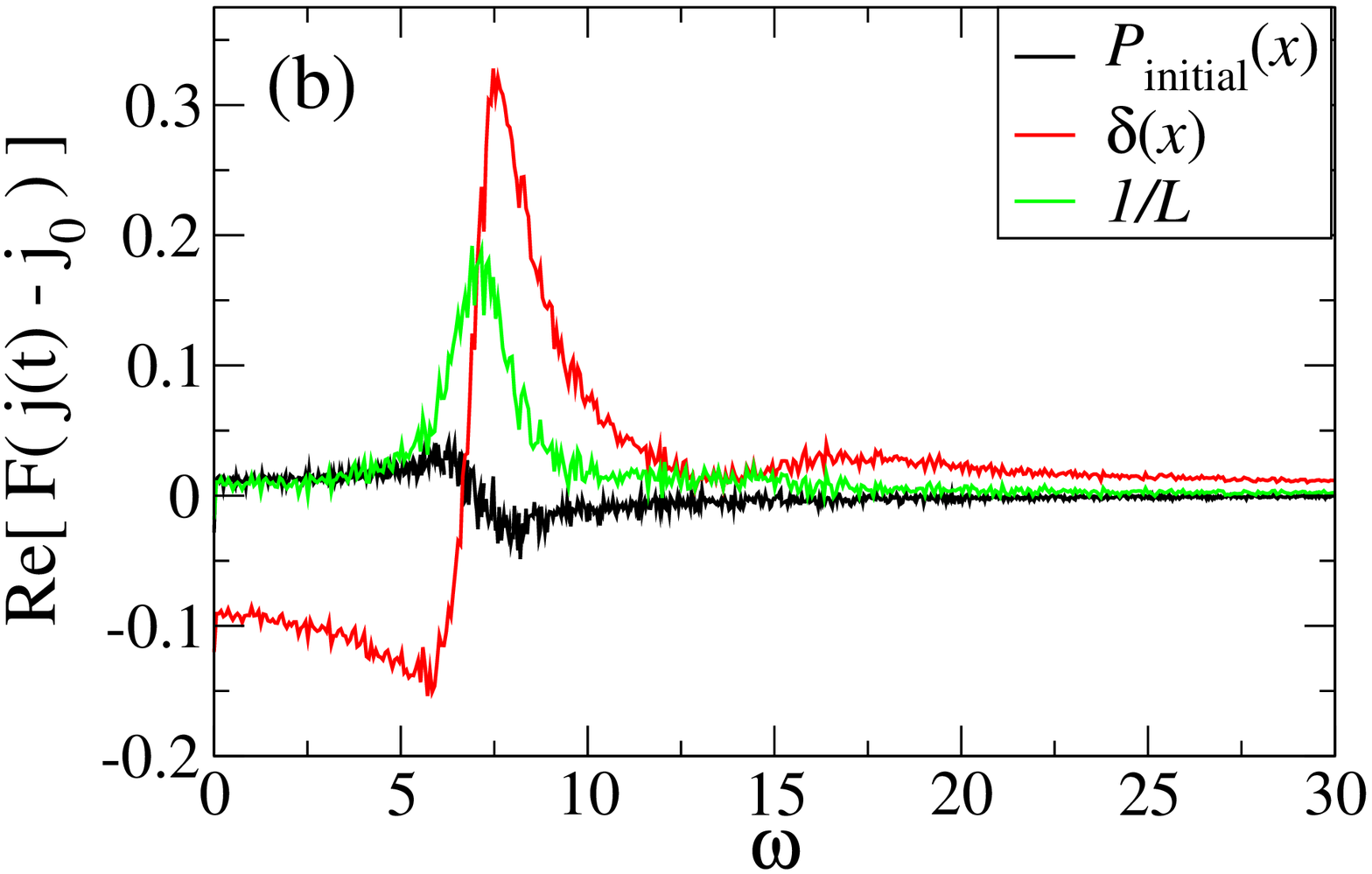}}
\end{center}
     \caption{(Color online)
     The Fourier transform of the time-dependent probability current towards the stationary state for (a) $F_0 = 0.90$ and (b) $ F_0 = 1.50$. In both cases we prepared the system at three different initial states:  (i) $P_{\mathrm{initial}} = P_0 (x) + \gamma P_0^\prime(x) $ with $\gamma = 0.08$ (black line), (ii) $ P_{\mathrm{initial}} =  \delta(x)$ (red line) and (iii) $P_{\mathrm{initial}} =  1/L$ (green line). None of the two latter initial conditions reproduces the behavior of response to the probability current. Nevertheless, in all cases the curves obtained show a resonance-like behavior near the location of the ``true resonance''. Thus, we can say that these experiments are able to qualitatively indicate  where the resonance peaks are probably located.  }
\label{fig:DIC.F0.90.F1.50}
\end{figure}
%
%
%

In Fig.~\ref{fig:RR.F0.98.F1.50} we test again the relaxation-response relation given by~\eqref{eq:RRrelation} for other values of the tilt. In Figs.~\ref{fig:RR.F0.98.F1.50}a and~\ref{fig:RR.F0.98.F1.50}b  we fix $F_0 = 0.98$ and  $F_0 = 1.50$ respectively. For such values of the tilt we measured the probability current $j_0 (F_0)$ for th unperturbed system giving $j_0 (0.98) = 0.28 \pm 0.005$ and  $j_0 (1.50) = 1.12 \pm 0.005$. According to Ref.~\cite{salgado-garcia} this gives the ``natural frequencies'' $\omega_0(0.98) \approx 1.74$ and $\omega_0(1.50) \approx 7.0$ respectively. We perturb the system with the same time-dependent forcing used in the experiment described above. We then calculated the mobility shown in the mentioned figures by means of the linear response to the perturbation (filled circles) and by means of the relaxation of the probability current (solid line). We observe again a good agreement of expression~\eqref{eq:RRrelation} in both numerical experiments. 

In the case $F_0 = 0.98$ it is interesting to notice that although the response is enhanced at the predicted frequency $\omega_0(0.98) \approx 1.74$, it does not seem to correspond to a maximum located at such frequency. Rather, we observe that the maximum seems to occur at $\omega = 0$. Actually,  in our simulation the response is measured for $\omega$'s bigger than one. However we are not able to observe any maximum in the interval $1 < \omega < 2$.  This could probably be due to either the lack of statistics or the fact that the imaginary part of the (non-vanishing) largest eigenvalue of such a system is actually smaller than the predicted in~\cite{salgado-garcia}. 

On the other hand, in the numerical experiments performed, we have seen that Eq.~\eqref{eq:RRrelation} is right within the accuracy of our simulations. The expression~\eqref{eq:RRrelation} states that the relaxation from a given initial state to the NESS is related to the response to the stationary PDF. The initial state referred above is very specific and it is interesting to see what happens with different initial conditions. In Fig.~\ref{fig:DIC.F0.90.F1.50}a and~\ref{fig:DIC.F0.90.F1.50}b we calculated the ``relaxation curves'' for the cases $F_0 = 0.90$ and $F_0 = 1.50$ respectively.  In both figure the initial conditions used were (i) $P_{\mathrm{initial}} = P_0 (x) + \gamma P_0^\prime(x) $ with $\gamma = 0.08$ (black line), (ii) $ P_{\mathrm{initial}} =  \delta(x)$ (red line) and (iii) $P_{\mathrm{initial}} =  1/L$ (green line). 
It is possible to observe that, although the initial conditions (ii) and (iii) do not reproduce the behavior of the response, they do indicate qualitatively the position of the resonance peaks. This fact is important since it can be used to predict the resonance character of a given system, not by performing directly the perturbation on it but observing its relaxation properties.


\section{DISCUSSION AND CONCLUSIONS.}
\label{sec:conclusion}


We have shown the existence of a resonant behavior that is exclusive of non-equilibrium stationary states. The requirement that the system be in a NESS is a necessary but not a sufficient condition for the resonant behavior be present. We proved the latter statement for systems with markovian dynamics which includes discrete and continuous systems. In particular, we have also found that the frequency-dependent mobility of a given system is related to the relaxation of the probability current to the stationary state  with a specific initial condition. We verified the presence of these resonances for the unperturbed system which consists of an ensemble of non-interacting overdamped particles in a tilted periodic potential. It has been shown that the correspondence established between the mobility and the Fourier transform of the relaxation of probability current (Eq.~\eqref{eq:RRrelation}) agreed with our numerical experiments for several values of the strength of tilt. Finally, we analyzed the relaxation properties of the system for other initial conditions different from the ``specific'' mentioned in the text. We found that the relaxation curves obtained do not reproduce the response of the system, but they do exhibit a ``peak'' near the resonant frequencies. On this basis, we say that it could be possible to predict the resonant behavior of a given system by analyzing its relaxation to the NESS.

We would like to remark that our results could be useful in understanding some phenomena like the relaxation properties of systems that attain a NESS~\cite{quian2010,ge2011non,asai2011,chou2009changing} or the so-called resonant transport~\cite{borromeo2008resonant,romanczuk2010quasideterministic}. It would also be interesting to see how the relaxation-response relation~\eqref{eq:RRrelation},  is connected to noise-enhanced stability of metastable states~\cite{fiasconaro2005signatures,dubkov2004noise,agudov2001noise}.

\section*{Acknowledges}

This work was supported by PROMEP through grant No. 103-5/11/4299. The author wishes to thank F. V\'azquez and  H. Larralde for carefully reading the manuscript and for giving useful comments on this work.

\bigskip

\appendix

\section{Detailed balance implies real eigenvalues}
\label{ape:DetBal-RealEigenVals}

Let the elements of $T$ be denoted, as usual, by $T_{i,j}$, with $i,j \in X$. The detailed balance condition states that $T_{i,j} \pi_i = T_{j,i} \pi_j$, where $\pi$ stands for the stationary probability vector for $T$. Following Ref.~\cite{levin2009markov}, we should notice that the latter can also be written as $\pi_j^{-1/2}T_{i,j} \pi_i^{1/2} = \pi_i^{-1/2} T_{j,i} \pi_j^{1/2}$. If we define $T^*_{i,j} := \pi_j^{-1/2}T_{i,j} \pi_i^{1/2}$ we can see that the matrix $T^*$ is symmetric and therefore its eigenvalues are real. Additionally, we can see that $T^*$ is obtained by a similarity transformation from $T$, since $T^* = S^{-1} T S$ where, $S_{i,j} = \pi_i^{1/2} \delta_{i,j}$. This implies that $T^*$ and $T$ have the same eigenvalues. Then it follows that if $T$ comply with the detailed balance condition, then it has only real eigenvalues. The proof of the latter statement for $\mathcal{L}$ as a Fokker-Planck operator is carried out in a similar way. First assume that $\mathcal{L}$ comply the detailed balance condition and define $P_0(x) := \langle x| P_0 \rangle$ as the corresponding equilibrium stationary state. Then note that the transformation defined by $\mathcal{S} = P_0(x)^{1/2} |x \rangle\langle x| $ has inverse  $\mathcal{S}^{-1} = P_0(x)^{-1/2} |x \rangle\langle x| $ since, 
\begin{eqnarray}
\mathcal{S}\mathcal{S}^{-1} &=& \bigg( P_0(y)^{1/2} |y \rangle \langle y| \bigg) \bigg( P_0(x)^{-1/2} |x \rangle \langle x| \bigg),
\nonumber 
\\
&=&
P_0(y )^{1/2}  P_0(x)^{-1/2} \delta(x,y) |y \rangle \langle x|.
\nonumber 
\\
&=&
\mbox{Id},
\nonumber 
\end{eqnarray}
where $\mbox{Id}$ stands for the identity operator. On the other hand we have that 
\begin{eqnarray}
\mathcal{S}^2 &=& \bigg( P_0(y)^{1/2} |y \rangle \langle y| \bigg) \bigg( P_0(x)^{1/2} |x \rangle \langle x| \bigg),
\nonumber 
\\
&=&
P_0(y )^{1/2}  P_0(x)^{1/2} \delta(x,y) |y \rangle \langle x|.
\nonumber 
\\
&=&
P_0(x) |x \rangle \langle x|.
\nonumber 
\end{eqnarray}
In Ref.~\cite{kurchan1998fluctuation} it has been shown that the detailed balance condition leads to the following symmetry property for the Fokker-Planck operator,
\begin{equation}
\mathcal{Q}^{-1}\mathcal{L} \mathcal{Q} = \mathcal{L}^\dagger, 
\label{eq:symmetry-kurchan}
\end{equation}
where $\mathcal{Q}$ is defined thorough the equation,
\[
\mathcal{Q} |x\rangle  =  P_0(x) |x \rangle.
\]
It is clear that a representation for $\mathcal{Q}$ is given by $\mathcal{S}^2$. This makes it possible rewrite Eq.~\eqref{eq:symmetry-kurchan} as,
\begin{eqnarray}
 \left( \mathcal{S}\mathcal{S}\right)^{-1} \mathcal{L} \mathcal{S}\mathcal{S} &=& \mathcal{L}^\dagger,
\nonumber  
\end{eqnarray}
or, equivalently, as,
\begin{eqnarray}
\mathcal{L}^* := \mathcal{S}^{-1} \mathcal{L} \mathcal{S} &=& \mathcal{S}\mathcal{L}^\dagger \mathcal{S}^{-1},
\nonumber  
\\
&=& 
\left( \mathcal{S}^{-1}\mathcal{L} \mathcal{S}\right)^\dagger = \left( \mathcal{L}^* \right)^\dagger .
\end{eqnarray}
The above equation states that, if the balance condition holds, there is a similarity transformation making the Fokker-Planck operator hermitian. Then, detailed balance condition implies that all the eigenvalues of $\mathcal{L}$ are real.
This proves that a system having a equilibrium state does not exhibit the resonant behavior predicted by~\eqref{eq:R-resonant}.

\section{Tilted periodic potentials at zero temperature}
\label{ape:zero-temp}

Consider an ensemble of non-interacting overdamped particles in a tilted periodic potential  $V(x) = V_p(x) - F_0 x$. Here $V_p(x)$ represents the periodic part of the potential, which has period $L$, and $F_0$ represent the tilt of the potential. The deterministic evolution equation for one overdamped particle in such a potential is given by,
\begin{equation}
\label{eq:unperturbed-deterministic}
\frac{dx}{dt} = f(x) = f_p(x) + F_0,
\end{equation}
where $f(x)$ is the gradient of the tilted periodic potential and $f_p(x) = -V_p^\prime(x)$ is the gradient of the corresponding periodic part. We will first consider the case in which $F_0$ exceeds the so called ``critical tilt'' $F_c := \max \{|f_p(x)|\}$. In such a case Eq.~\eqref{eq:unperturbed-deterministic} have only running solutions and, as we anticipated in Section~\ref{sec:tilted-potentials} it exhibits a resonant behavior in the frequency-dependent mobility. 

Let $ x (t)$ be a solution of~\eqref{eq:unperturbed-deterministic} with a given initial condition $x_0$ and consider the perturbed system
\begin{equation}
\label{eq:perturbed-deterministic}
\frac{dy}{dt} = f(y) + \varepsilon F(t)
\end{equation}
where $\varepsilon \ll 1$ is a dimensionless control parameter and $F(t)$ is a time-dependent periodic external forcing with period $T$. For $\varepsilon$ small we can measure the response of the system as the deviation of the perturbed trajectory $y(t)$  from the unperturbed one $ x(t)$ having the same initial condition $y(0)= y_0 = x_0$. Then we will analyze the the behavior of the ``linear response'' $\eta (t) := (y(t) -x(t))/\varepsilon $. From the above we can see that the evolution equation for $\eta(t)$ is approximately given by,
\begin{equation}
\label{eq:linear-deterministic}
\frac{d\eta}{dt} = f^\prime(x(t)) \eta + F(t),
\end{equation}
at first order in $\varepsilon$. The above equation can be easily solved if we notice that the integration factor $\nu(t)$ is given by,
\begin{equation}
\label{eq:int-factor}
\nu(t) := \exp\left(-\int_0^t f^\prime(x(s))ds \right).
\end{equation} 
This lets us write a solution for $\eta(t)$ as follows,
\[
\eta(t) = \frac{1}{\nu(t)} \int_0^t \nu(s) F(s) ds 
\] 
The integration factor $\nu$ can be written in a more compact form if we use the fact that the solution $x(s)$ is always increasing and therefore invertible. A change of variables in the integration in Eq.~\eqref{eq:int-factor} shows that,
\[
\nu(t) = \frac{f(x_0)}{f\left ( x(t) \right) },
\]
which can be used to express $\eta$ as,
\begin{equation}
\label{eq:eta-sol}
\eta(t) = f(x(t)) \int_0^t\frac{F(s)}{f(x(s)) } ds.
\end{equation}

In Appendix~\ref{ape:running-solution} we show that any running solution $x(t)$ of the unperturbed system with $F_0 > F_c $ has the property $x(t + \tau) = x(t) + L $, where $\tau$ is the time that the particle takes to travel one spatial period. From this property it follows that $f(x(s))$ is periodic with ``natural frequency''  $\omega_0 = 2\pi /\tau$, which is proportional to the mean velocity $\overline{v} = L/\tau$. If we denote by $c_n$ the coefficients of the Fourier series of $1/f(x(s))$, and take $F(s) = e^{i\omega t}$, we obtain for $\eta$,
\begin{equation}
\eta(t) = f(x(t)) \sum_{n} c_n \frac{ e^{i(\omega-n\omega_0) t}-1}{ \omega-n\omega_0 }.
\end{equation}
From this we can see that the deviation of the perturbed trajectories from the unperturbed ones is enhanced (and even diverges) when the frequency of the driving force is $2 \pi /L$ times the mean velocity of the particle,
\begin{equation}
\label{eq:freq-det}
\omega_n = \frac{2 \pi \overline{v}}{ L}
\end{equation}

To tackle the case $F_0 < F_c$ we should notice that Eq.~\eqref{eq:linear-deterministic} and its solution~\eqref{eq:eta-sol} is valid in this case. The only difference is that $x(t)$ has no longer the property $x(t+\tau) = x(t)+L$. Instead, the solution $x(t)$ to the differential equation~\eqref{eq:unperturbed-deterministic} has the asymptotic behavior $x(t) = x^* + C_0 \exp(-\lambda t)$ for $t\to \infty$. Here $x^*$ is a root of $f(x)$, $\lambda = f^\prime (x^*)$ and $C = x_0 - x^* $. Thus, for large $t$ we can expand $f(x(s))$ and $f(x(t))$ around $x^*$ to obtain for $\eta$,
\[
\eta(t) = \int_0^t e^{-\lambda (t-s)} F(s)ds.
\] 
To obtain the linear response to the particle current we fix $F(t) = e^{i\omega y}$ and take the time derivative to $\eta$ giving,
\[
\dot \eta(t) = \left(  1 - \frac{i\omega}{ \lambda + i \omega} \right) e^{-\lambda t} + \frac{i\omega e^{i\omega t}}{\lambda + i \omega} .
\] 
From the latter we see that $\dot \eta $ is asymptotically periodic, and the amplitude of (the real part of) such a function is given by,
\[
A (\omega) = \frac{\omega^2}{ \lambda^2 + \omega^2},
\]
which is a monotonically increasing function of $\omega$.

\section{Analytic running solutions.}
\label{ape:running-solution}

The evolution equation for overdamped particles in a tilted periodic potential $V(x) = V_p - x F_0$ is 
\begin{equation}
   \frac{dx}{dt} = f(x)
\end{equation}
where $f(x) = f_p(x)  + F_0 $ is the gradient of $V(x)$. Since it is assumed that $F_0$ is greater than the critical tilt $F_c := \max \{|f_p(x)|\}$, we have that $f(x)$ has no real roots. This means that we can integrate the above equation with the initial value $x(0)=x_0$, giving, 
\begin{equation}
\label{eq:ape:sol1}
t = \int_{x_0}^{x(t)} \frac{dy}{f(y)}.
\end{equation}

Define $\tau$ as the time that the particle takes to travel one spatial period, i.e.,
\[
\tau := \int_{0}^{L} \frac{dx}{f(x)},
\]
and notice that this time does not depend on where the particle starts its motion. Now define the invertible function
\[
h(x) := \frac{L }{\tau } \int_{0}^{x} \frac{dy}{f(y)}.
\]
Due to periodicity of $f(x)$ we have that $h(x+L) = h(x) + L$. In terms of this function we can write Eq.~\eqref{eq:ape:sol1} as,
\[
\frac{Lt}{\tau} = h(x(t)) - h(x_0).
\]
The invertibility of $h$ lets us write $x(t)$ as,
\begin{equation}
\label{eq:solution}
x(t) = h^{-1} \left( h(x_0) + \frac{Lt}{\tau} \right).
\end{equation}
It is easy to see that $h^{-1}$ has also the property $h^{-1}(x+L) = h^{-1}(x) + L$, from which we obtain that $x(t+ \tau) = x(t) + L$.

\nocite{*}

\bibliography{ResResp.ref.2011}

\end{document}